\documentclass{article}
\topskip=\baselineskip
\let\ep=\epsilon

\let\la=\langle
\let\ra=\rangle

\let\e=\emph

\let\ct=\cite
\let\lf=\left
\let\rt=\right
\let\bv=\mathbf

\let\dt=\cdot
\let\del=\nabla

\let\q=\widehat

\let\h=\hbar
\let\rta=\rightarrow
\let\dy=\displaystyle
\let\ty=\textstyle

\let\x=\times
\let\hl=\hfill
\newcommand{\m}{\mbox}
\newcommand{\ol}[1]{\makebox[\textwidth][s]{#1}}

\newcommand{\eqdf}{\stackrel{\mathrm{def}}{=}}
\newcommand{\hf}{\ensuremath{{\scriptstyle\frac{1}{2}}}}
\newcommand{\thf}{\ensuremath{{\scriptstyle\frac{3}{2}}}}

\newcommand{\be}{\begin{equation}}
\newcommand{\ee}{\end{equation}}
\newcommand{\dd}[3]{\\ \m{}\\ \ol{\m{#1}\hl\m{${\dy #2}$}\hl\m{#3}}\\ \m{}\\}
\newcommand{\re}[2]{\dd{}{#1}{(#2)}}

\newcommand{\ba}{\begin{array}}
\newcommand{\ea}{\end{array}}
\newcommand{\bea}{\begin{eqnarray}}
\newcommand{\eea}{\end{eqnarray}}
\newcommand{\beas}{\begin{eqnarray*}}
\newcommand{\eeas}{\end{eqnarray*}}

\newcommand{\qH}{\q{H}}

\newcommand{\vu}{\bv{u}}
\newcommand{\vv}{\bv{v}}
\newcommand{\vp}{\bv{p}}

\newcommand{\vE}{\bv{E}}
\newcommand{\vB}{\bv{B}}
\newcommand{\vj}{\bv{j}}
\newcommand{\vA}{\bv{A}}
\newcommand{\vPsi}{\bv{\Psi}}

\newcommand{\vk}{\bv{k}}
\newcommand{\vr}{\bv{r}}

\newcommand{\vz}{\bv{0}}

\newcommand{\qvp}{\q{\vp}}

\newcommand{\wop}{\la\vk|\psi_1(t)\ra}
\newcommand{\wtp}{\la\vk|\psi_2(t)\ra}
\newcommand{\wom}{\la\psi_1(t)|-\vk\ra}
\newcommand{\wtm}{\la\psi_2(t)|-\vk\ra}
\title{Equivalence of Maxwell's source-free equations to the \\
       time-dependent Schr\"{o}dinger equation for a solitary \\
       particle with two polarizations and Hamiltonian $|c\qvp|$}
\author{Steven Kenneth Kauffmann \\
        American Physical Society Senior Life Member}
\date{43 Bedok Road \\
      {\#}01-11 \\
      Country Park Condominium \\
      Singapore 469564 \\
      Handphone: +65 9370 6583 \\
      \m{} \\
      and \\
      \m{} \\
      Unit 802, Reflection on the Sea \\
      120 Marine Parade \\
      Coolangatta QLD 4225 \\
      Australia \\
      Tel/FAX: +61 7 5536 7235 \\
      Mobile:  +61 4 0567 9058 \\
      \m{} \\
      Email: SKKauffmann@gmail.com}
\begin{document}
\maketitle
\begin{abstract}
It was pointed out in a previous paper that although neither the Klein-Gordon
equation nor the Dirac Hamiltonian produces sound solitary free-particle
relativistic quantum mechanics, the natural square-root relativistic 
Hamiltonian for a nonzero-mass free particle does achieve this.  Failures of
the Klein-Gordon and Dirac theories are reviewed: the solitary Dirac free
particle has, \e{inter alia}, an invariant speed well in excess of $c$ and
staggering spontaneous Compton acceleration, but no pathologies whatsoever
arise from the square-root relativistic Hamiltonian.  Dirac's key
misapprehension of the underlying four-vector character of the time-dependent,
configuration-representation Schr\"{o}dinger equation for a solitary particle
is laid bare, as is the invalidity of the standard ``proof'' that the
nonrelativistic limit of the Dirac equation is the Pauli equation.  Lorentz
boosts from the particle rest frame point uniquely to the square-root
Hamiltonian, but these don't exist for a massless particle.  Instead, Maxwell's
equations are dissected in spatial Fourier transform to separate nondynamical
longitudinal from dynamical transverse field degrees of freedom.  Upon their
decoupling in the absence of sources, the transverse field components are seen
to obey two identical time-dependent Schr\"{o}dinger equations (owing to two
linear polarizations), which have the massless free-particle diagonalized
square-root Hamiltonian.  Those fields are readily \e{modified} to conform to
the attributes of solitary-photon \e{wave functions}.  The wave functions'
relations to the \e{potentials} in radiation gauge are also  worked out.  The
exercise is then repeated \e{without} the considerable benefit of the spatial
Fourier transform.
\end{abstract}

\subsection*{Introduction}

It was pointed out in a previous paper~\ct{Ka1} that a \e{solitary}
free relativistic nonzero-mass particle is described \e{without any
pathology whatsoever} by the \e{natural} time-dependent
Schr\"{o}dinger equation,
\re{
i\h\partial(|\psi(t)\rangle)/\partial t = \sqrt{m^2c^4 + |c\qvp |^2}\:
 |\psi(t)\rangle,
}{1}
whereas the widely used relativistic free-particle Klein-Gordon and
Dirac equations are generally acknowledged not to be up to this
simple task~\ct{B-D}.  For example, negative energy solutions of
the Klein-Gordon equation \e{fail} to be orthogonal to their
\e{positive} energy counterparts that have the \e{same momentum}.
This contradicts a fundamental property of quantum theory that
makes its probability interpretation possible; unsurprisingly it
is well-known that Klein-Gordon theory can yield \e{negative}
probabilities~\ct{B-D}.  This particular pathology of the
second-order in time Klein-Gordon equation cannot arise if the
solitary free particle is described by a standard first-order
in time Schr\"{o}dinger equation with a Hermitian Hamiltonian
operator, such as that of Eq.~(1).  The \e{particular} Hamiltonian
operator of Eq.~(1), namely $\sqrt{m^2c^4 + |c\qvp |^2}$,
has the \e{additional} virtue of \e{fully adhering} to the classical
Correspondence Principle, being that it is the \e{direct}
quantization of the \e{correct classical Hamiltonian} for a
solitary relativistic free particle of mass $m$.  It is to be
noted in particular that solitary free \e{relativistic}
particles \e{cannot} have negative energies if solitary free
\e{nonrelativistic} particles are to be restricted to having
only \e{nonnegative kinetic energies}!

In light of the problems the second-order in time Klein-Gordon
equation has in describing the solitary relativistic free particle,
Dirac appreciated the need for elementary \e{relativistic}
quantum mechanics to \e{return} to the standard first-order in
time Schr\"{o}dinger equation format with Hermitian Hamiltonian
operator that serves elementary \e{nonrelativistic} solitary-particle
quantum mechanics so admirably.  Lamentably, however, Dirac
was less responsive to the exacting requirements of the classical
Correspondence Principle than he was, like Klein, Gordon and
Schr\"{o}dinger, misdirectedly concerned about the fact that the
solitary free particle Hamiltonian operator
$\sqrt{m^2c^4 + |c\qvp |^2}$ turns out to be a \e{nonlocal} entity
in configuration representation: it doesn't seem to have occurred
to these pioneers that this fact in \e{no} way stymies the fruitful
application of \e{perturbation approximations}---the relativistic
\e{corrections} to the atomic physics in which they
were interested are obviously \e{very well-suited} to this approach,
being \e{compatibly small}.  Dirac unfortunately
\e{rejected} the Correspondence Principle appropriate square-root
Hamiltonian operator of Eq.~(1) in favor of a \e{misconceived
linearization} of it in terms of the components of the momentum
operator $\qvp$ and the mass $m$, for which he argued on the basis
of a \e{fundamental misapprehension} of the manner in which the 
\e{solitary-particle} time-dependent Schr\"{o}dinger equation in
configuration representation,
\[i\h\partial(\la\vr|\psi(t)\ra)/\partial t = \la\vr|\qH|\psi(t)\ra,\]
is related to the \e{covariance requirements of special relativity}~%
\ct{Sch, Dir, B-D}.  It is clear that the operator $\partial/\partial t$
on the \e{left-hand side} of this equation is the \e{time component}
of the \e{four-vector operator} $c\partial/\partial x_{\mu}$, and
the operator $\qH$ on the \e{right-hand side} of this equation is the
\e{time component} of the \e{four-vector operator} $c\q{p^{\mu}}$, where
$\q{p^{\mu}}\eqdf(\qH/c,\,\qvp)$.  Moreover, it was \e{postulated} by
Schr\"{o}dinger, and is a \e{basic consequence} of Dirac's \e{own}
canonical commutation rule, that,
\[-i\h\del_{\vr}(\la\vr|\psi(t)\ra) = \la\vr|\qvp|\psi(t)\ra,\]
so that the \e{full four-vector equation},
\[i\h\partial(\la\vr|\psi(t)\ra)/\partial x_{\mu} =
  \la\vr|\q{p^{\mu}}|\psi(t)\ra,\]
is \e{guaranteed to hold} in solitary-particle quantum mechanics!
Since the operator $\partial/\partial x_{\mu}$ is \e{patently} a
\e{Lorentz covariant} four-vector, the covariance requirements
of special relativity are met in solitary-particle quantum
mechanics by simply requiring that the Hamiltonian operator
$\qH$ be selected so as to ensure that the four-vector operator
$\q{p^{\mu}} = (\qH/c,\,\qvp)$ \e{also} transforms between
inertial frames as a \e{Lorentz covariant} four-vector.
This requirement is \e{automatically fulfilled} by
\e{scrupulous adherence to the strongest possible form of the
classical Correspondence Principle}, i.e., that $\qH$ be the
\e{quantization} of precisely that \e{classical Hamiltonian} H
which has been carefully checked to be appropriate to fully
relativistic solitary-particle \e{classical} mechanics!  For 
the \e{free} solitary particle of nonzero mass m, this physically
\e{methodical} and \e{highly conservative} approach leaves us
with \e{no option} but to \e{accept} Eq.~(1) as its \e{correct}
time-dependent Schr\"{o}dinger equation description!  This
\e{even} extends to \e{free} spin $\hf$ particles of nonzero
mass: notwithstanding that spin $\hf$ itself is a nonclassical
attribute, the nonrelativistic Pauli Hamiltonian for such
a particle \e{automatically reduces} to the \e{usual}
nonrelativistic \e{purely kinetic-energy Hamiltonian} in
the \e{free-particle} limit, and one can \e{always} find
an inertial frame of reference in which a \e{free particle
of nonzero mass} is \e{completely nonrelativistic}!

Dirac, however, was much too focused on trying to cobble up a
relativistic solitary-particle Hamiltonian operator which is
\e{local in configuration representation} to be in any frame
of mind to appreciate this \e{profound link} between the
strongest form of the classical Correspondence Principle and
the \e{requirement of Lorentz covariance} in solitary-particle
\e{quantum mechanics}.  Instead of pondering the \e{details}
of \e{how} the requirement of Lorentz covariance actually
impacts the time-dependent solitary-particle Schr\"{o}dinger
equation in configuration representation, Dirac was content
to assume that relativistic covariance merely requires that
there be essentially complete symmetry in the formal treatment
of space and time coordinates~\ct{Sch, B-D, Dir}.  As a result,
he \e{completely missed the point} that the time-dependent
Schr\"{o}dinger equation relates the \e{time derivative}
operator to an \e{energy} operator, \e{neither of which} are
Lorentz \e{scalars}, but \e{each of which} is the \e{time
component} of a Lorentz \e{four-vector}.  Not having \e{assimilated}
these basics, he conjured from whole cloth a \e{nonexistent
Lorentz scalar} which he perceived this equation to \e{split into
two nonscalar fragments for the express purpose of displaying
the fragment which is proportional to the time derivative on
the left-hand side of the equality sign}!  Thus primed, Dirac
``concluded'' that his phantom scalar's ``completion
fragment'', which is everything on the \e{right-hand side}
of the equality sign, ``must'' therefore be \e{linear in the
space gradient}, which suited his purpose \e{perfectly}, as it
results in a \e{local} ``Hamiltonian operator'' in configuration
representation!  Following this ``eureka moment'', which was the
\e{fruit} of his \e{mistakenly identifying as a scalar the time
component of a four-vector}, Dirac \e{failed to reflect} on
whether a Hamiltonian operator that is \e{linear} in the space
gradient, and thus in the momentum, could \e{really} be
relativistically correct in light of the firmly established
understanding that a solitary free particle's Hamiltonian is
\e{ineluctably} the \e{time-component} of a \e{Lorentz-covariant
four-vector} whose \e{remaining three components} are $c$ times
that free particle's three-momentum!  This understanding,
\e{conjoined with the Lorentz transformation itself}, in fact
\e{determines} that the \e{square-root} Hamiltonian which occurs
in Eq.~(1) is the \e{only correct one} for the nonzero-mass free
particle!

Dirac \e{also} paid no heed to the fact that a solitary free-particle
Hamiltonian operator which is \e{linear} in the space gradient,
and therefore in the momentum operator, has, in light of
Heisenberg's equation of motion, the \e{unavoidable consequence}
that the free particle's \e{velocity} is \e{completely
independent} of its \e{momentum}, which is an \e{astounding
contradiction of nonrelativistic free-particle physics},
quantum or classical!  Dirac determined the \e{coefficients}
of his misconceived linearized Hamiltonian operator by
requiring that its \e{square} be equal to the \e{square}
of the square-root Hamiltonian operator of Eq.~(1),
which is a perilously weakened \e{surrogate} for the
strong classical Correspondence Principle that \e{produces} the
square-root Hamiltonian operator of Eq.~(1) in the first place!
It results in these coefficients satisfying the well-known Dirac-matrix
\e{anticommutation relations}~\ct{B-D}.  The free-particle
\e{velocity operator}, which involves \e{only} these coefficients,
is thereupon determined to equal the speed of light $c$ times a
three-vector of the Dirac matrices, which each square to unity.
Therefore the \e{speed} of \e{any} free Dirac particle turns out
to have the universal \e{superluminal} value $\sqrt{3}\,c$,
\e{irrespective of its momentum}!  The free-particle Dirac equation
in fact yields \e{more} such inordinately \e{unphysical} results.
Upon using the misconceived linearized Dirac Hamiltonian operator
in conjunction with Heisenberg's equation of motion to calculate
the free particle's spontaneous \e{acceleration}, one finds that
its \e{magnitude} has a \e{minimum} value of order of the ``Compton
acceleration'' $mc^3/\h$, which, for the electron, is about
$10^{28}g$, an \e{absolutely staggering} violation of Newton's
First Law of Motion for a free particle!  The natural square-root
Hamiltonian operator of Eq.~(1) gives \e{nil} spontaneous acceler%
ation, a result that is, of course, in \e{complete agreement} with
Newton's First Law of Motion for a free particle.  It as well gives
the \e{correct expression} for the relativistic free particle's
\e{velocity} in terms of its momentum.  The \e{extreme disparity}
of the results of the natural square-root Hamiltonian operator
versus those of the \e{misbegotten linearized} Dirac Hamiltonian
operator is an overwhelming object lesson on the dangers inherent
in \e{any weakening} of the strongest sensible version
of the classical Correspondence Principle.

Although it is routinely claimed that the Dirac equation reduces
to the nonrelativistic Pauli equation for a spin $\hf$ particle 
when the particle's momentum magnitude is much less than
$mc$~\ct{B-D}, the ``proof'' of this assertion is definitely
\e{invalid}, being \e{unsalvageably dependent} on the lapse
of \e{forgetting} that at $\vp = \vz$ the two \e{lower
components} of the Dirac spinor in the \e{standard representation}
have the time-dependence phase factor $e^{+(imc^2/\h)t}$, which is
\e{totally different} from the analogous factor $e^{-(imc^2/\h)t}$
that occurs in this spinor's two \e{upper components}!  Nor does
this assertion remotely accord with some of the most elementary
``physics'' consequences of the \e{free} Dirac and Pauli theories
at \e{vanishing momentum}.  The latter's Hamiltonian operator is
just the nonrelativistic kinetic energy operator $|\qvp|^2/(2m)$,
and its particle speed operator is, of course, $|\qvp|/m$.  Thus a
\e{free} Pauli particle eigenstate of vanishing momentum has vanishing
speed.  For the \e{free} Dirac theory, we have \e{already seen} that
the particle speed operator is \e{even simpler}, namely the \e{identity
operator} times the \e{universal superluminal speed} $\sqrt{3}\,c\,$!
So a \e{free} Dirac particle eigenstate of vanishing momentum \e{still}
has this problematic extreme speed!

Furthermore, \e{notwithstanding its spin $\hf$ attribute}, the
\e{free} Pauli particle's \e{orbital angular momentum is exactly
conserved}, i.e., the rate of change of its orbital angular momentum
\e{vanishes identically}.  The \e{free} Dirac particle's orbital
angular momentum, however, is coupled with \e{astonishing strength}
to its spin: as the free Dirac particle's \e{momentum magnitude}
tends toward \e{zero}, the \e{dimensionless ratio} of the
\e{magnitude} of the rate of change of its orbital angular momentum
to its kinetic energy \e{increases monotonically without bound},
beginning from the \e{asymptotic} ultrarelativistic \e{dimensionless
ratio value} $\sqrt{2}\,$!  In other words, \e{far} from having the
\e{exactly conserved orbital angular momentum of the free Pauli
particle}, the free Dirac particle's spin-orbit \e{torque magnitude}
always \e{well exceeds} that particle's \e{own kinetic energy}, and
the dimensionless ratio of these two quantities in fact becomes
\e{arbitrarily} large at \e{low enough} particle momentum!

All of these \e{stunningly unphysical} properties of
the Klein-Gordon and Dirac equations with regard to
the description of a solitary relativistic free particle
\e{notwithstanding}, and \e{even in spite} of the fact
that the simple square-root Hamiltonian operator of
Eq.~(1)---which is the \e{unique consequence} of the
classical Correspondence Principle for a solitary relativistic
free particle---doesn't  partake of \e{any} such pathologies,
it \e{still} has always been the Dirac and Klein-Gordon equations,
\e{rather} than Eq.~(1), that are inducted into relativistic
quantum field theory.  The reason for this, of course, is that 
\e{antiparticles} were first observed sometime \e{after} Dirac
began to speculate about mechanisms which could serve to \e{hide}
the physically problematic \e{unbounded-below} negative energy
spectra that are a feature of his and the Klein-Gordon equations,
but which simply \e{do not occur} for Eq.~(1).  Had Dirac \e{not}
been so speculating, the existence of antiparticles would have
been regarded as an \e{energy degeneracy} of nature's full field
theoretic Hamiltonian, and an explanation for that \e{degeneracy}
would have been sought, following the grand tradition established
by Wigner, Weyl and others, entirely in terms of the effect of a
\e{symmetry} possessed by that full Hamiltonian.  A particle and
its antiparticle are \e{distinguishable}, and each can have only
\e{positive} energy, so it is \e{entirely natural} that they
should be described by \e{two entirely independent quantum fields,
with each having purely positive energy}.  In other words, had
the Klein-Gordon and Dirac equations, with their problematic
unbounded-below negative energy spectra \e{never been concocted},
it would have been perfectly straightforward to accommodate the
discovery of antiparticles in a simple, logical framework that is
very strongly grounded in physical precedent.  The straightforward
use of the \e{purely positive energy} Eq.~(1) in conjunction with
\e{symmetry} postulates to accommodate \e{antiparticles} has the
\e{theoretical advantage} that it \e{as well} automatically
accommodates a sensible theory of a solitary free relativistic
particle, which the Dirac and Klein-Gordon equations are
\e{utterly unsuited} to do.  There is no physical reason whatsoever
that \e{nonrelativistic} solitary particle theory should not
link to relativistic particle physics in a completely \e{smooth
fashion}, which is what Eq.~(1) transparently enables.  Furthermore,
both the Klein-Gordon and Dirac equations historically arose as
\e{eccentric offshoots} of Eq.~(1), motivated \e{not} by legitimate
physics concerns, but by an \e{irrational distaste} for the
\e{nonlocal} character of Eq.~(1) in \e{configuration representation}.
This means that the Klein-Gordon and Dirac equations were \e{not
designed ab initio} to accommodate \e{both} a particle and its
antiparticle: this is a role into which historical happenstance
has \e{pushed} them---by their \e{actual patrimony} they were
\e{designed} to accommodate \e{only a single type of particle}!
Nowadays, it is known that particle-antiparticle symmetry can
be slightly \e{broken}, as CP noninvariance experiments have
shown (given the dominance of particles over antiparticles in
our immediate surroundings, it would be \e{astonishing} if
particle-antiparticle symmetry were \e{not} in fact broken).
But the Dirac and Klein-Gordon fields, \e{not} having been
\e{designed to accommodate two particles}, are highly stressed
to accommodate two slightly \e{nondegenerate} particles, which
is what corresponds to the existent symmetry breaking.  It is
obvious that the model with two \e{independent} positive-energy
fields for particle and antiparticle offers \e{vastly more
flexibility to accommodate symmetry breaking} than do the
claustrophobic Dirac and Klein-Gordon models, which shoehorn
two particles into a field structure that was \e{designed to
accommodate just one}.  As one example, two independent fields
easily accommodate two slightly different masses: there is
simply no way to have a single Dirac or Klein-Gordon field with
more than one mass.

For a nonzero-mass solitary free particle, the relativistic 
square-root Hamiltonian operator of Eq.~(1) is \e{completely
determined} by the Lorentz transformation.  This is because
one can \e{always} find an inertial frame in which a
\e{solitary} free particle of mass $m$ is at rest, i.e., has
four-momentum $(mc, \vz)$.  The \e{Lorentz transformation} to
the inertial frame in which this particle has velocity $\vv$,
where $|\vv|<c$, then takes the particle's four-momentum to,
\[ (mc(1-|\vv|^2/c^2)^{-\hf},\: m\vv(1-|\vv|^2/c^2)^{-\hf}) =
   (E(\vv)/c,\: \vp(\vv)),\]
which, together with the \e{identity},
\[ mc^2(1-|\vv|^2/c^2)^{-\hf} =
 \sqrt{m^2c^4 + |cm\vv|^2(1 -|\vv|^2/c^2)^{-1}},\]
implies that,
\[E(\vv) = \sqrt{m^2c^4 + |c\vp(\vv)|^2}.\]
Since the classical precursor of the square-root Hamiltonian
operator for the solitary free-particle of mass $m$ that occurs
in Eq.~(1) is thus \e{mandated by the very nature of the Lorentz
transformation}, it is little wonder that Dirac's misconceived
effort to \e{linearize} the square-root character of this
Hamiltonian operator has consequences which \e{terribly violate}
well-known relativistic properties of a free particle: we have
seen that these consequences include the blatantly unphysical
universal \e{superluminal} free particle speed $\sqrt{3}\,c$
\e{irrespective} of the particle momentum, a \e{minimum}
spontaneous free-particle acceleration magnitude of order of the
Compton acceleration $mc^3/\h$, namely about $10^{28}g$ for the
electron, which staggeringly violates Newton's First Law of
Motion for a free particle, and the gross failure to \e{conserve}
free-particle orbital angular momentum, which the nonrelativistic 
spin $\hf$ Pauli theory free particle definitely \e{does}.

For a \e{zero-mass free particle}, however, there is \e{no}
inertial frame in which that particle is at rest, so we
cannot readily derive its Hamiltonian from the Lorentz
transformation, as we have done for the nonzero-mass free
particle. Confirmation that the Hamiltonian operator given
by Eq.~(1) continues to be correct for a \e{massless} solitary
free particle must be sought elsewhere.  We therefore turn to
the study of electromagnetic radiation, which is supposed
to consist of \e{massless photons}.  Surprisingly, we shall
see that Maxwell's \e{classical} equations for \e{pure}
electromagnetic radiation can be recast into a form that is
in essence that of the time-dependent Schr\"{o}dinger equation
of Eq.~(1) with $m=0$.  Because of the particle's vanishing
mass, Planck's constant $\h$ can be factored out of \e{both
sides} of Eq.~(1), since $\qvp = -i\h\del_{\vr}$ in configuration
representation and $\qvp = \h\vk$ in Fourier vector variable
$\vk$-representation.  That Planck's constant \e{drops out of the
relativistic solitary free-particle time-dependent Schr\"{o}dinger
equation in the $m=0$ case} is a key factor in allowing that equation
to be related to the putatively ``classical'' Maxwell theory.

To make further progress, we must \e{dissect} Maxwell's four
equations themselves: these are a mixed bag of dynamical field
equations of motion and nondynamical constraint conditions.
Since the time-dependent Schr\"{o}dinger equation is \e{purely
dynamical} in character, it will be necessary to properly
resolve the nondynamical constraint conditions, a task which
we now undertake.

\subsection*{The electromagnetic field as a constrained dynamical system}

Since any time-dependent Schr\"{o}dinger equation is linear and
\e{homogeneous}, only the \e{source-free} (i.e., pure radiation)
version of Maxwell's equations could possibly correspond to such
an equation.  But the resolution of the two \e{nondynamical
constraints} amongst the the four Maxwell equations can be carried
out even in the \e{presence} of the source terms, so we shall
initially retain those terms.  The four Maxwell equations for
the electromagnetic field $(\vE, \vB)$ with four-current source
$(\rho, \vj/c)$ are comprised of Coulomb's law,
\re{\del\dt\vE = \rho,
}{2a}
Faraday's law,
\re{\del\x\vE = -\dot{\vB}/c,
}{2b}
Gauss' law,
\re{\del\dt\vB = 0,
}{2c}
and Maxwell's law,
\re{\del\x\vB = (\vj + \dot{\vE})/c,
}{2d}
which, together with Coulomb's law, implies the current
conservation condition,
\re{\del\dt\vj + \dot{\rho} = 0.
}{2e}
Coulomb's and Gauss' laws both involve \e{no} time derivatives
of the electromagnetic field, so they are in the nature of
\e{nondynamical constraints} on that field, whereas Faraday's
and Maxwell's law's, which both \e{do} involve first time
derivatives of the electromagnetic field, have the character
of dynamical equations of motion of that field.  If one is
presented with a set of $N$ variables which are subject to both
nondynamical equations of constraint and dynamical equations of
motion, it is standard practice to search for $N$ \e{functions}
of those $N$ variables with the property that all the dynamical
equations of motion involve \e{only} a subset of $N-k$ of these
functions, while all the nondynamical equations of constraint
involve \e{only} the \e{remaining} subset of $k$ functions.  The
\e{first} set of $N-k$ functions is not subject to \e{any}
nondynamical equations of constraint (these apply \e{exclusively}
to the \e{second} set of $k$ functions), and are regarded as a set
of purely \e{dynamical variables} for the system.  The \e{second}
set of $k$ functions, to which \e{no} dynamical equations of motion
apply, may analogously be regarded as a set of purely
\e{nondynamical variables} for the system.  The equations of
motion satisfied by the $N-k$ \e{unconstrained} dynamical variables
are then typically summarized by means of a Lagrangian or
Hamiltonian from which they follow, respectively, via the
Euler-Lagrange or classical Hamiltonian equations equations of
motion.  Hamiltonization of such a maximal set of unconstrained
dynamical variables opens the way to the system's \e{quantization}
via either the Hamiltonian phase-space path integral~\ct{Ka2}, or,
equivalently, the slightly strengthened self-consistent extension
of Dirac's canonical commutation rule~\ct{Ka3}.

For the electromagnetic \e{field}, $N$ is formally infinite, but
we can still usefully discuss the number of \e{field} degrees
of freedom; e.g., the electromagnetic field $(\vE(\vr, t),
\vB(\vr, t))$ has six \e{field} degrees of freedom.
\e{Both} the equations of motion and those of constraint are
\e{linear} for the electromagnetic field, so one can expect
the extraction of a maximal subset of unconstrained dynamical
variables (actually unconstrained dynamical \e{fields}) to
involve appropriate \e{linear transformations} of components
of the electromagnetic field $(\vE(\vr, t), \vB(\vr, t))$.
Furthermore, consideration of the Coulomb and Gauss equations
of constraint quickly makes it clear that $\del\dt\vE(\vr, t)$
and $\del\dt\vB(\vr, t)$ (which vanishes identically!) are each
purely nondynamical \e{single} field degrees of freedom, and that
\e{no additional} purely \e{nondynamical} field degrees of freedom
are available to be extracted from the six field degrees of
freedom of the electromagnetic field system $(\vE(\vr, t),
\vB(\vr, t))$.  Therefore the electromagnetic field must have
\e{four} unconstrained, purely \e{dynamical} field degrees of
freedom.  To cleanly separate the nondynamical $\del\dt\vE(\vr, t)$
and $\del\dt\vB(\vr, t)$ from the purely dynamical part of
$(\vE(\vr, t), \vB(\vr, t))$, a \e{hypothetical coordinate
system} in which one of the three components of the electric
field $\vE(\vr, t)$ \e{is} just $\del\dt\vE(\vr, t)$ and also
in which one of the three components of the magnetic field
$\vB(\vr, t)$ \e{is} just $\del\dt\vB(\vr, t)$ would be very
convenient.  In such a hypothetical coordinate system, the set
of the remaining two components of $\vE(\vr, t)$, together with
the remaining two components of $\vB(\vr, t)$, would comprise
the four unconstrained, purely \e{dynamical} electromagnetic
field degrees of freedom.  It turns out to be technically 
most straightforward to precede attempted \e{implementation}
of this sort of idea by \e{spatial Fourier transformation} of
the electromagnetic field $(\vE(\vr, t), \vB(\vr, t))$ and its
four-current source $(\rho(\vr, t), \vj(\vr,t)/c)$.  We define,
\re{(\vE(\vk, t), \vB(\vk, t))\eqdf 
(2\pi)^{-\thf}{\ty\int} d^3\vr\,e^{-i\vk\dt\vr}(\vE(\vr, t), \vB(\vr, t)),
}{3a}
which is the ``unitary'' Fourier transform.  Also,
\re{(\rho(\vk, t), \vj(\vk, t)/c)\eqdf 
(2\pi)^{-\thf}{\ty\int} d^3\vr\,e^{-i\vk\dt\vr}(\rho(\vr, t), \vj(\vr, t)/c).
}{3b}
In the penultimate section of this paper, we shall essay
the trickier task of attempting to reveal the dynamical
time-dependent Schr\"{o}dinger equation character of Maxwell's
source-free equations directly in configuration representation,
without resort to this spatial Fourier transformation.
It is worth remarking at this stage that since the charge
density $\rho(\vr, t)$ is a real-valued function, and
the \e{same} is true of all the Cartesian components of
$(\vE(\vr, t)$, $\vB(\vr, t))$ and $\vj(\vr,t)$, the
corresponding spatial Fourier transforms of all these
entities have the property that their complex
conjugation is equivalent to \e{reversing the sign}
of their \e{Fourier vector argument} $\vk$.  Some
key manipulations that are carried out further on rely
heavily on this technical point.  The spatial Fourier
transformation of $\del\dt\vE(\vr, t)$ comes out be 
$i\vk\dt\vE(\vk, t)$, which, in a \e{coordinate system} 
that has $\vu_L(\vk)\eqdf\vk/|\vk|$ as \e{one} of its
three orthogonal unit vectors, is equal to $i|\vk|$ times
the $\vu_L(\vk)$-component of $\vE(\vk, t)$, which we denote
as $E_L(\vk, t)$.  Coulomb's law thus obviously implies the
nondynamical equation,
\re{ E_L(\vk, t) = -i\rho(\vk, t)/|\vk|,
}{4a}
and, analogously, Gauss' law implies the nondynamical equation
\re{ B_L(\vk, t) = 0.
}{4b}
We can therefore be quite confident that $E_L(\vk, t)$ and
$B_L(\vk, t)$ exhaust the nondynamical components of
$\vE(\vk, t)$ and $\vB(\vk, t)$ respectively, and that the
\e{remaining} two components of each of these two fields will
be purely dynamical, i.e., free of any nondynamical constraint.
But to demonstrate this in detail, we must explicitly display
the \e{remaining two} mutually orthogonal unit vectors, which
are \e{each} as well orthogonal to $\vu_L(\vk)$, and then work
out the consequences of the Maxwell equations for the remaining
two components of both $\vE(\vk, t)$ and $\vB(\vk, t))$ in
\e{that} coordinate system, in order to verify that \e{purely
dynamical equations of motion} which involve \e{only} these
four components result.  In the particular case that the
four-current source $(\rho(\vk, t), \vj(\vk,t)/c)$ \e{vanishes},
we \e{also} need to demonstrate that the now \e{homogeneous}
equations of motion obtained for these four unconstrained
\e{dynamical components} of $(\vE(\vk, t), \vB(\vk, t))$ are
\e{equivalent} to the schematic Schr\"{o}dinger Eq.~(1) with
$m=0$---note as well that in this source-free case the two
\e{nondynamical components} $E_L(\vk, t)$ and $B_L(\vk, t)$
of $(\vE(\vk, t), \vB(\vk, t))$ \e{vanish identically}, as
is seen from Eqs.~(4).

In order to obtain two mutually orthogonal unit vectors which
are both \e{also} orthogonal to the unit vector $\vu_L(\vk) =
\vk/|\vk|$, we display $\vu_L(\vk)$ in \e{Cartesian coordinates}:
it is simply the well-known unit vector in the \e{radial} direction
that the Fourier vector argument $\vk$ points toward, expressed in
terms of that vector's spherical polar angles $\phi_\vk$ and
$\theta_\vk$,
\re{\vu_L(\vk)=(\cos\phi_\vk\sin\theta_\vk, \sin\phi_\vk\sin\theta_\vk,
   \cos\theta_\vk).
}{5a}
Now because $\vu_L(\vk)=\vk/|\vk|$, $\vu_L(-\vk)=-\vu_L(\vk)$, i.e.,
$\vu_L(\vk)$ has the same \e{odd parity} that $\vk$ has.  Therefore
the parity flip mapping $\vk\rta -\vk$ corresponds to the polar
angular mapping $\theta_\vk\rta\theta_\vk + \pi$, because this sends
$\sin\theta_\vk\rta -\sin\theta_\vk$ and $\cos\theta_\vk\rta -\cos
\theta_\vk$, thus sending, from Eq.~(5a), $\vu_L(\vk)\rta -\vu_L(\vk)$.
However, if we instead choose to carry out the polar angular mapping
$\theta_\vk\rta\theta_\vk + \pi/2$, then $\sin\theta_\vk\rta
\cos\theta_\vk$, $\cos\theta_\vk\rta -\sin\theta_\vk$, and
$\vu_L(\vk)\rta\vu_1(\vk)$, where,
\re{\vu_1(\vk)\eqdf (\cos\phi_\vk\cos\theta_\vk, \sin\phi_\vk\cos\theta_\vk,
   -\sin\theta_\vk),
}{5b}
is readily checked to be a \e{unit vector} that is \e{orthogonal} to
$\vu_L(\vk)$.  The parity flip angular mapping
$\theta_\vk\rta\theta_\vk + \pi$ reveals that $\vu_1(\vk)$ is \e{also}
of \e{odd} parity.  With the mutually orthogonal unit vectors
$\vu_L(\vk)$ and $\vu_1(\vk)$ in hand, we can now readily construct a
\e{third} unit vector $\vu_2(\vk)$ which is orthogonal to \e{both} of
these,
\re{\vu_2(\vk)\eqdf\vu_L(\vk)\x\vu_1(\vk)=(-\sin\phi_\vk, \cos\phi_\vk, 0).
}{5c}
It is immediately seen that $\vu_2(\vk)$ is of \e{even} parity. By using
the identity $\bv{a}\x(\bv{b}\x\bv{c}) = \bv{b}(\bv{a}\dt\bv{c}) -
\bv{c}(\bv{a}\dt\bv{b})$, or, alternatively, the spherical polar angular 
representations given by Eqs.~(5a)--(5c), it is readily checked that
$\vu_1(\vk)$, $\vu_2(\vk)$ and $\vu_L(\vk)$ comprise a ``right-handed''
orthonormal local vector triad, i.e.,
\re{\vu_1(\vk)\x\vu_2(\vk)=\vu_L(\vk),\m{\ }\vu_2(\vk)\x\vu_L(\vk)=\vu_1(\vk)
\m{\ and\ } \vu_L(\vk)\x\vu_1(\vk)=\vu_2(\vk).
}{5d}
\indent
Turning now to the implications of Maxwell's equations in this
coordinate system, we have already noted that the Coulomb and
Gauss laws imply the two nondynamical Eqs.~(4a) and (4b).  Upon
spatial Fourier transformation, Faraday's law, Eq.~(2b), becomes,
\re{i\vk\x\vE(\vk, t) = -\dot\vB(\vk,t)/c.
}{6a}
Noting that $\vk = |\vk|\vu_L(\vk)$, and that,
\[\vE(\vk, t)=E_1(\vk, t)\vu_1(\vk)+E_2(\vk, t)\vu_2(\vk)
  +E_L(\vk, t)\vu_L(\vk),\]
where,
\[E_1(\vk, t)\eqdf\vu_1(\vk)\dt\vE(\vk, t),\m{\ }
  E_2(\vk, t)\eqdf\vu_2(\vk)\dt\vE(\vk, t)\m{\ and\ }
  E_L(\vk, t)\eqdf\vu_L(\vk)\dt\vE(\vk, t),\]
and analogously for $\dot\vB(\vk,t)$, for which the
Gauss law result embodied by Eq.~(4b)
\e{already} permits us to conclude that $\dot B_L(\vk, t)=0$,
we apply Eq.~(5d) to the left-hand side of Eq.~(6a),
and thereby obtain the two \e{additional} equations,
\re{i\dot B_1(\vk, t)=-|c\vk|E_2(\vk, t),
}{6b}
and,
\re{i\dot B_2(\vk, t)=|c\vk|E_1(\vk, t).
}{6c}
Before we turn to Maxwell's law, Eq.~(2d), it
is convenient to treat the current conservation
condition, Eq.~(2e), which is a \e{constraint
on the four-current source} that follows from
Maxwell's and Coulomb's laws.  Upon spatial
Fourier transformation, Eq.~(2e) becomes,
\re{i\vk\dt\vj(\vk, t)=-\dot\rho(\vk, t),
}{6d}
which immediately yields the longitudinal
source current component in terms of the
rate of change of the charge density,
\re{j_L(\vk, t)=i\dot\rho(\vk, t)/|\vk|.
}{6e}
Upon spatial Fourier transformation, Maxwell's
law, Eq.~(2d), becomes,
\re{i\vk\x\vB(\vk, t)=(\vj(\vk, t) + \dot\vE(\vk, t))/c.
}{6f}
The left-hand side of Eq.~(6f) has a vanishing component
in the $\vu_L(\vk)$-direction, and the joint consequence
of that and the Coulomb law result embodied by Eq.~(4a)
for its right-hand side is simply the constraint on
the longitudinal source current component that is embodied
by eq.~(6e).  More interesting are the two equations that
follow from the components of Eq.~(6f) in the $\vu_1(\vk)$
and $\vu_2(\vk)$ directions---these bear a strong resemblance
to the Eqs.~(6b) and (6c) which follow from Faraday's law,
\re{i\dot E_1(\vk, t)=|c\vk|B_2(\vk, t)-ij_1(\vk, t),
}{6g}
and,
\re{i\dot E_2(\vk, t)=-|c\vk|B_1(\vk, t)-ij_2(\vk, t).
}{6h}
Aside from the purely \e{source} constraint requirement
of Eq.~(6e) and the reconfirmation that $\dot B_L(\vk, t)$
must vanish, which is already a consequence of Eq.~(4b)
(which is itself the result of the Gauss law), the Faraday
and Maxwell laws have yielded four \e{dynamical} equations
of motion, namely Eqs.~(6b), (6c), (6g) and (6h), which
involve \e{only} the four \e{transverse} field components
$E_1(\vk, t)$, $B_2(\vk, t)$, $E_2(\vk, t)$ and $B_1(\vk, t)$.
Absolutely \e{no} nondynamical equations of constraint for
\e{any} of these four transverse field components have
eventuated from any of the Maxwell equations.  It is
therefore clear that the six field degrees of freedom of
$(\vE(\vk, t), \vB(\vk, t))$ have now been \e{successfully
partitioned} into \e{four} unconstrained, purely dynamical
\e{transverse} field degrees of freedom and \e{two} purely
nondynamical \e{longitudinal} field degrees of freedom
$E_L(\vk, t)$ and $B_L(\vk, t)$, whose values are actually
\e{given} by the simple nondynamical constraints of Eqs.~(4a)
and (4b).  In \e{addition}, it has, of course, transpired that
the four-current source $(\rho(\vk, t),\vj(\vk, t)/c)$
\e{cannot} be chosen \e{arbitrarily}, but is subject to the
\e{source constraint} given by Eq.~(6e).

\subsection*{Linear algebraic decoupling of the four
             transverse dynamical fields}

Eq.~(6g) for the dynamical transverse fields $E_1(\vk, t)$
and $B_2(\vk, t)$ is clearly coupled to Eq.~(6c), and
likewise Eq.~(6h) for the dynamical transverse fields
$E_2(\vk, t)$ and $B_1(\vk, t)$ is clearly coupled to
Eq.~(6b).  Some investigators may be tempted to
decouple these equations by taking \e{second} time
derivatives, but such an approach is \e{entirely
unnecessary} and involves a risk of \e{introducing}
extraneous solutions that \e{don't} actually apply
to these equations---indeed taking an unwarranted
second time derivative is precisely how the
unphysical, unbounded-below negative energy
spectrum was inadvertently forced into the
relativistic Klein-Gordon equation for a
nonzero-mass free particle.  Eqs.~(6g) and
(6c) are easily decoupled by the straightforward
expedient of taking their sum and difference, and
the same applies to Eqs.~(6h) and (6b).  Adding
Eq.~(6c) to Eq.~(6g) yields,
\re{i\partial(E_1(\vk, t) + B_2(\vk, t))/\partial t =
    |c\vk|(E_1(\vk, t) + B_2(\vk, t)) - ij_1(\vk, t),
}{7a}
while subtracting Eq.~(6b) from Eq.~(6h) yields,
\re{i\partial(E_2(\vk, t) - B_1(\vk, t))/\partial t =
    |c\vk|(E_2(\vk, t) - B_1(\vk, t)) - ij_2(\vk, t).
}{7b}
One can also subtract Eq.~(6c) from Eq.~(6g) to obtain,
\re{i\partial(E_1(\vk, t) - B_2(\vk, t))/\partial t =
    -|c\vk|(E_1(\vk, t) - B_2(\vk, t)) - ij_1(\vk, t),
}{7c}
and add Eq.~(6b) to Eq.~(6h) to obtain,
\re{i\partial(E_2(\vk, t) + B_1(\vk, t))/\partial t =
    -|c\vk|(E_2(\vk, t) + B_1(\vk, t)) - ij_2(\vk, t).
}{7d}
Now it turns out that Eq.~(7c) is \e{not independent}
of Eq.~(7a); in fact, Eq.~(7c) is actually \e{equivalent}
to Eq.~(7a)!  The reason for this is somewhat involved; it
is related to the previously mentioned fact that for any
Cartesian component of $\vE(\vk, t)$, $\vB(\vk, t)$,
or $\vj(\vk, t)$, complex conjugation is equivalent
to changing the \e{sign} of the Fourier vector argument
$\vk$.  Making matters a bit more complicated is the
fact that $E_1(\vk, t)$ and $E_2(\vk, t)$ are \e{not
Cartesian components} of $\vE(\vk, t)$ because
$E_1(\vk, t) = \vE(\vk, t)\dt\vu_1(\vk)$ and
$E_2(\vk, t) = \vE(\vk, t)\dt\vu_2(\vk)$.  Because
$\vu_1(\vk)$ is of \e{odd parity} in its argument
$\vk$, complex conjugation of $E_1(\vk, t)$ is
equivalent to changing \e{both} the sign of its
\e{argument} $\vk$ \e{and} its \e{overall} sign!
However, because $\vu_2(\vk)$ is of \e{even parity}
in its argument $\vk$, complex conjugation of
$E_2(\vk, t)$ is equivalent to merely changing the
sign of its \e{argument} $\vk$.  Exactly the same
distinction with regard to complex conjugation holds
between $B_1(\vk, t)$ and $B_2(\vk, t)$, and as well
between $j_1(\vk, t)$ and $j_2(\vk, t)$.  Now if we
take the complex conjugate of both sides of Eq.~(7c)
and apply what we have just learned, the result is,
\re{-i\partial(-E_1(-\vk, t) - B_2(-\vk, t))/\partial t =
    -|c\vk|(-E_1(-\vk, t) - B_2(-\vk, t)) - ij_1(-\vk, t).
}{7e}
Upon combining signs in Eq.~(7e), we find that it resembles
Eq.~(7a) in every respect, except for the fact that all
occurrences of the Fourier vector \e{argument} $\vk$
have effectively had their sign reversed.  However, because
Eq.~(7e) is supposed to hold irrespective of \e{what} value
is assumed by $\vk$, we are free to make the simple one-to-one
formal transformation $\vk\rta -\vk$ which turns Eq.~(7e) into
Eq.~(7a).  Furthermore, if we take the complex conjugate of
both sides of Eq.~(7d) and apply to it what we have learned
above, the result is,
\re{-i\partial(E_2(-\vk, t) - B_1(-\vk, t))/\partial t =
    -|c\vk|(E_2(-\vk, t) - B_1(-\vk, t)) + ij_2(-\vk, t).
}{7f}
Upon reversing the sign of both sides of Eq.~(7f), we find
that it resembles Eq.~(7b) in every respect, except for the
fact that all occurrences of the Fourier vector \e{argument}
$\vk$ have effectively had their sign reversed.  But we are,
of course, again justified in making the simple one-to-one
formal transformation $\vk\rta -\vk$ which turns Eq.~(7f)
into Eq.~(7b).

We have thus succeeded in replacing the \e{four coupled}
equations of motion for the dynamical transverse fields
by \e{two} nontrivially complex-valued and \e{fully
decoupled} such equations, namely Eqs.~(7a) and (7b).
If we multiply both of these equations through by $\h$,
and then set both of the transverse source currents
$j_1(\vk, t)$ and $j_2(\vk, t)$ to zero, Eqs.~(7a) and
(7b) assume precisely the schematic form of Eq.~(1)
with $m=0$, i.e., they are of the form of time-dependent
Schr\"{o}dinger equations for a solitary relativistic
massless free particle.  The fact that there are \e{two}
such equations suggests, in light of the detailed
electromagnetic field composition of each of their two
apparent ``wave functions'', that they describe the
amplitudes for two \e{linear polarization} states of the
solitary massless particle.  We shall now further
investigate this interesting source-free limit of
Maxwell's equations.

\subsection*{The Schr\"{o}dinger character of the source-free
             Maxwell equations}

When the four-current source $(\rho(\vk, t), \vj(\vk, t)/c)$
vanishes altogether, Eqs.~(4) show that the two nondynamical
longitudinal electromagnetic field elements $E_L(\vk, t)$ and
$B_L(\vk, t)$ vanish identically as well.  The \e{only} physics
that remains is \e{purely dynamical and transverse}, and is
fully describled by the two relativistic, massless, solitary
free-particle Schr\"{o}dinger-style equations,
\re{i\h\partial(E_1(\vk, t) + B_2(\vk, t))/\partial t =
    |c\h\vk|(E_1(\vk, t) + B_2(\vk, t)),
}{8a}
and,
\re{i\h\partial(E_2(\vk, t) - B_1(\vk, t))/\partial t =
    |c\h\vk|(E_2(\vk, t) - B_1(\vk, t)),
}{8b}
which follow from Eqs.~(7a) and (7b) in the source-free
situation.  The detailed structure of the two putative
``wave functions'' in terms of the transverse electromagnetic
field components strongly suggests that they represent the
amplitudes for the two possible transverse linear polarization
states of the solitary, massless, free electromagnetic field
particle.  There is a technical snag, however, which bars
such an interpretation from being immediately made: the
``wave functions'' that appear in the Schr\"{o}dinger equations
of Eqs.~(8) are sums and differences of \e{transverse electromagnetic
field components}, which have the character of \e{energy-density
amplitudes}, whereas \e{true solitary-particle wave functions} have
the character of \e{probability-density amplitudes}.  To get a
feeling for just \e{what} energy is represented by the two complex-%
valued ``wave functions'' of Eqs.~(8), we wish to integrate the sum of
their absolute squares over all of the Fourier vector-variable
$\vk$-space.  We begin by integrating over just the absolute
square of the ``wave function'' of Eq.~(8a),
\re{ {\ty\int d^3\vk(|E_1 + B_2|^2 =
    \int d^3\vk(|E_1|^2 + |B_2|^2)
    +\int d^3\vk(E_1^\ast B_2 + B_2^\ast E_1)},
}{9a}
where we have temporarily suppressed writing out the \e{arguments}
of the transverse field components to save space.  However, bearing
in mind the discussion between Eqs.~(7d) and (7e), we have
that, $(E_1(\vk, t))^\ast = -E_1(-\vk, t)$ and
$B_2(\vk, t)=(B_2(-\vk, t))^\ast$, from which we readily deduce
that $\int d^3\vk\,E_1^\ast B_2 = -\int d^3\vk\,B_2^\ast E_1$,
and therefore that the second integral on the right-hand side of
Eq.~(9a) vanishes.  Analogous arguments show that when one
integrates over the absolute square of the ``wave function'' of
Eq.~(8b), the integration over the corresponding two cross
terms vanishes as well.  Therefore, the result of integrating
over the \e{sum} of the absolute squares of these two
``wave functions'' is,
\re{ {\ty\int d^3\vk(|E_1 + B_2|^2 + |E_2 - B_1|^2) =
    \int d^3\vk(|E_1|^2 + |E_2|^2 + |B_1|^2 +|B_2|^2)}.
}{9b}
Now the integral on the right-hand side of Eq.~(9b) is
equal to twice the total energy present in the transverse
electromagnetic field components, which is, of course twice
the total electromagnetic field energy, since the
longitudinal components of the electromagnetic field vanish
identically in the source-free case.  Therefore the energy
represented by the two complex-valued ``wave functions'' of
Eqs.~(8) is twice the total electromagnetic field energy.

Now let us suppose that the \e{sole source} of the
electromagnetic field energy which is present is a \e{solitary}
photon.  That photon's energy density in $\vk$-space (which is
effectively momentum-space, since $\h\vk$ \e{is} the photon's
momentum) is then equal to \e{half} of the sum of the absolute
squares of the two complex-valued ``wave functions'' of Eq.~(8),
as we have learned from Eq.~(9b).  Now insofar as the solitary
photon \e{has} its Fourier vector variable equal to $\vk$, i.e.,
insofar as it \e{has} momentum $\h\vk$, it clearly \e{also has
energy} $|c\h\vk|$.  Therefore, we can \e{convert} our photon's
\e{energy density} in $\vk$-space---which is \e{half} of the sum
of the squares of the two complex-valued ``wave functions'' of
Eqs.~(8)---to its \e{probability density} in $\vk$-space by simply
dividing that energy density by $|c\h\vk|$.  This implies that we
can \e{convert} each of the two transverse electromagnetic field
``wave functions'' of Eqs.~(8) to a proper \e{solitary photon}
wave function (whose absolute square yields a \e{probability
density}) by dividing it by $(2|c\h\vk|)^{\hf}$.  It is clear
that both of these proper \e{solitary photon wave functions} will
satisfy the \e{very same Schr\"{o}dinger equation} that the two
transverse electromagnetic field component ``wave functions'' of
Eqs.~(8) satisfy: the factor of $(2|c\h\vk|)^{-\hf}$ \e{doesn't
interfere} with the validity of that time-dependent Schr\"{o}dinger
equation.  Therefore, when only a solitary photon is present, its
two linear polarization wave function components (complex-valued
probability amplitudes) are given in terms of the corresponding
transverse electromagnetic field components by,
\re{ \wop = (2|c\h\vk|)^{-\hf}(E_1(\vk, t) + B_2(\vk, t)),
}{10a}
and,
\re{ \wtp = (2|c\h\vk|)^{-\hf}(E_2(\vk, t) - B_1(\vk, t)).
}{10b}
It is convenient to as well explicitly write down the
parity-reversed complex conjugates of these solitary-photon
linear polarization wave function components,
\re{ \wom = -(2|c\h\vk|)^{-\hf}(E_1(\vk, t) - B_2(\vk, t)),
}{10c}
and,
\re{ \wtm = (2|c\h\vk|)^{-\hf}(E_2(\vk, t) + B_1(\vk, t)),
}{10d}
because, with these in hand, the relationships of the solitary-%
photon linear polarization wave function components to the
transverse electromagnetic field components can be \e{inverted},
\re{ E_1(\vk, t) = (|c\h\vk|/2)^{+\hf}(\wop - \wom),
}{10e}
\m{}
\re{ E_2(\vk, t) = (|c\h\vk|/2)^{+\hf}(\wtp + \wtm),
}{10f}
\m{}
\re{ B_1(\vk, t) = -(|c\h\vk|/2)^{+\hf}(\wtp - \wtm),
}{10g}
\m{}
\re{ B_2(\vk, t) = (|c\h\vk|/2)^{+\hf}(\wop + \wom).
}{10h}
It is worth explicitly reiterating that the two complex-%
valued linear polarization components of the solitary-photon
wave function satisfy the massless case of the relativistic
free-particle time-dependent Schr\"{o}dinger equation that
is given by Eq.~(1),
\re{i\h\partial(\wop)/\partial t = |c\h\vk|\,\wop,
}{11a}
\m{}
\re{i\h\partial(\wtp)/\partial t = |c\h\vk|\,\wtp.
}{11b}
Finally, it is worthwhile to relate the solitary free-%
photon's complex-valued wave function to the components of
the electromagnetic \e{four-vector potential} to which it
corresponds.  The electromagnetic four-vector potential
does have a gauge ambiguity issue which unfortunately is
not fully resolved by the relativistically invariant
Lorentz condition---suppression of the ensuing timelike
and longitudinal ``ghost radiation''~\ct{Gu, Bl} requires
a further stipulation: probably the most intuitively
appealing is to require the scalar potential to be
\e{uniquely determined}, in strictly \e{homogeneous}
and \e{causal} fashion, by the charge density, which is,
after all, its notional \e{source} after imposition of
the Lorentz condition.  This produces results that
are no less definite than those of the Coulomb gauge%
---in fact these two gauges produce \e{identical}
results for all \e{static} charge densities---but
without the Coulomb gauge's disconcerting instantaneous
scalar potential response at arbitrarily large
distances to charge density \e{change}.  In the
present source-free case, both gauges are, in fact,
identical to the \e{radiation gauge}~\ct{BD2},
$\del\dt\vA(\vr, t) = \phi(\vr, t) = 0$, which causes
the four-vector potential to have \e{only four transverse
dynamical field degrees of freedom}, in complete
\e{agreement} with the situation discussed above for the
\e{electromagnetic field} in this source-free case.
The relation of the electromagnetic field to the
four-vector potential is, of course, given by,
\re{\vB(\vr, t) = \del\x\vA(\vr, t),
}{12a}
and,
\re{\vE(\vr, t) = -\del\phi(\vr, t) - \dot\vA(\vr, t)/c,
}{12b}
which, in spatial Fourier transform become,
\re{\vB(\vk, t) = i\vk\x\vA(\vk, t),
}{12c}
and,
\re{\vE(\vk, t) = -i\vk\phi(\vk, t) - \dot\vA(\vk, t)/c.
}{12d}
Upon applying to it Eq.~(5d), Eq.~(12c) readily yields
the two transverse components of $\vA(\vk, t)$ in terms
of those of $\vB(\vk, t)$,
\re{A_1(\vk, t) = -iB_2(\vk, t)/|\vk|,
}{13a}
\m{}
\re{A_2(\vk, t) = iB_1(\vk, t)/|\vk|,
}{13b}
and Eq.~(12d) \e{immediately} yields the two transverse
components of $\dot\vA(\vk, t)$ in terms of those of
$\vE(\vk, t)$,
\re{\dot A_1(\vk, t) = -cE_1(\vk, t),
}{13c}
\m{}
\re{\dot A_2(\vk, t) = -cE_2(\vk, t),
}{13d}
Upon putting Eqs.~(10e) through (10h) into Eqs.~(13a)
through (13d) above, we obtain,
\re{ A_1(\vk, t) = -i(|c\h\vk|/2)^{+\hf}(\wop + \wom)/|\vk|,
}{14a}
\m{}
\re{ A_2(\vk, t) = -i(|c\h\vk|/2)^{+\hf}(\wtp - \wtm)/|\vk|,
}{14b}
\m{}
\re{\dot A_1(\vk, t) = -c(|c\h\vk|/2)^{+\hf}(\wop - \wom),
}{14c}
\m{}
\re{\dot A_2(\vk, t) = -c(|c\h\vk|/2)^{+\hf}(\wtp + \wtm).
}{14d}
Eqs.~(14a) through (14d) can now be inverted,
\re{ \wop = (2|c\h\vk|)^{-\hf}(i|\vk|A_1(\vk, t) - \dot A_1(\vk, t)/c),
}{15a}
\m{}
\re{ \wtp = (2|c\h\vk|)^{-\hf}(i|\vk|A_2(\vk, t) - \dot A_2(\vk, t)/c),
}{15b}
\m{}
\re{ \wom = (2|c\h\vk|)^{-\hf}(i|\vk|A_1(\vk, t) + \dot A_1(\vk, t)/c),
}{15c}
\m{}
\re{ \wtm = -(2|c\h\vk|)^{-\hf}(i|\vk|A_2(\vk, t) + \dot A_2(\vk, t)/c).
}{15d}
From Eqs.~(15a) and (15b) it is apparent that the correct
Schr\"{o}dinger equation quantization of the solitary free
photon requires \e{not only} the two transverse components
of $\vA(\vk, t)$, but \e{as well} the two transverse
components of $\dot\vA(\vk, t)$.  The two linear polarization
state wave function components, $\wop$ and $\wtp$, are \e{each}
ineluctably complex-valued objects in a way that is thoroughly
\e{nonsuperficial}: it requires \e{two} ``classical'' field
degrees of freedom, such as \e{both} $E_1(\vk, t)$ and
$B_2(\vk, t)$, or \e{both} $A_1(\vk, t)$ and $\dot A_1(\vk, t)$,
to comprise \e{one} such deeply complex-valued quantum
wave function component.  Of course this bodes well for the
\e{next} level of quantization, wherein our Schr\"{o}dinger
equation wave function components are themselves promoted
to become operators which have prescribed commutation
relations with their own Hermitian conjugates: this reflects
their complex-valued makeup from \e{independent fields} which
are interpreted as being \e{mutually canonically conjugate},
a status for which the pair $A_1(\vk, t)$ and $\dot A_1(\vk, t)$
are, of course, prime candidates.  We see that the automatic
solitary photon ``first quantization'' that is simply part and
parcel of the very \e{nature} of Maxwell's supposedly ``classical''
equations also \e{automatically} has properties which
\e{anticipate} and \e{facilitate} ``second quantization''.
Once \e{transverse source currents} are present, the solitary
photon Schr\"{o}dinger equation becomes \e{inhomogeneous}, i.e.,
it no longer \e{is} a Schr\"{o}dinger equation, as we clearly
see from Eqs.~(7a) and (7b).  The \e{inhomogeneity} of what,
in the source-free case, had been the solitary photon
Schr\"{o}dinger equation, of course bespeaks \e{the creation
and destruction} of such photons.  It is quite remarkable,
however, just how well-organized the solitary photon wave function
is \e{ab initio} for rising to the challenges of the eventually
necessary ``second quantization''.

\subsection*{Configuration space approach to the Schr\"{o}dinger
character of Maxwell's source-free equations}

Having gained insight from dissection of the Maxwell equations in
spatially Fourier-transformed formulation, we now try our hand
at teasing out the Schr\"{o}dinger character of their transverse
dynamical segment directly in configuration representation.  We
have learned that ``dynamical'' and ``transverse'' are effectively
synonyms for Maxwell's equations, so we simply focus on
separating the electromagnetic fields $\vE(\vr, t)$ and $\vB(\vr, t)$,
and also the current $\vj(\vr, t)$, into their physically natural
transverse and longitudinal parts.  This, of course, requires no
action whatsoever for $\vB(\vr, t)$, which Gauss' law, Eq.~(2c),
marks a purely transverse.  The longitudinal part of $\vE(\vr, t)$
is the negative of the gradient of the very same scalar potential
$\phi(\vr, t)$ which describes $\vE(\vr, t)$ fully in the electro%
static limit, i.e.,
\re{
\vE_L(\vr, t) = -\del\phi(\vr, t),
}{16a}
where, of course,
\re{
\phi(\vr, t)\eqdf (4\pi)^{-1}{\ty\int}d^3\vr'\,\rho(\vr', t)/|\vr - \vr'|.
}{16b}
Because of the Green's function identity,
\[\del_{\vr}^2(1/|\vr -\vr'|) = -4\pi\,\delta^{(3)}\:(\vr -\vr'),\]
we have that,
\re{
\del^2\phi(\vr, t) = -\rho(\vr, t),
}{16c}
and,
\re{
\del\dt\vE(\vr, t) = \del\dt\vE_L(\vr, t) = \rho(\vr, t),
}{16d}
as required by the Coulomb law, Eq.~(2a).  Eq.~(16d) follows from
Eq.~(16a), Eq.~(16c) and,
\re{
\vE(\vr, t) = \vE_L(\vr, t) + \vE_T(\vr, t),
}{16e}
the separation of $\vE(\vr, t)$ into its longitudinal and transverse
parts, where, of course, by definition,
\re{
\del\dt\vE_T(\vr, t) = 0\ \m{and}\ \del\x\vE_L(\vr, t) = \vz.
}{16f}
Now the current conservation condition given by Eq.~(2e) has a
formal structure that is very similar to that of Coulomb's law,
and therefore permits an analogous determination of $\vj_L(\vr, t)$,
the longitudinal part of the current $\vj(\vr, t)$,
\re{
\vj_L(\vr, t) = \del\dot\phi(\vr, t),
}{16g}
which, together with Eq.~(16c) and the relations for $\vj(\vr, t)$,
$\vj_L(\vr, t)$ and $\vj_T(\vr, t)$ which are analogous to Eqs.~(16e)
and (16f), implies that,
\re{
\del\dt\vj(\vr, t) = \del\dt\vj_L(\vr, t) = -\dot\rho(\vr, t),
}{16h}
as required by the current conservation condition of Eq.~(2e).

We are now in a position to reexpress the Maxwell and Faraday laws
in terms of \e{only} the \e{transverse} fields $\vE_T(\vr, t)$ and
$\vB(\vr, t)$, and the transverse current $\vj_T(\vr, t)$.  In
particular, Eqs.~(16g) and (16a) permit us to deduce that,
\[\vj(\vr, t) + \dot\vE(\vr, t) = \vj_T(\vr, t) + \dot\vE_T(\vr, t),\]
which permits the Maxwell law of Eq.~(2d) to be rewritten,
\re{
\del\x\vB = (\vj_T + \dot\vE_T)/c.
}{17a}
From Eqs.~(16a) and (16e), or from Eq.~(16f), we can deduce that
$\del\x\vE(\vr, t) = \del\x\vE_T(\vr, t)$, which permits the Faraday
law of Eq.~(2b) to be rewritten,
\re{
\del\x\vE_T =  -\dot\vB/c.
}{17b}
This completes the divorce of the dynamical Maxwell and Faraday
laws from the purely nondynamical variable $\vE_L(\vr, t)$,
whose value is given in detail by Eqs.~(16a) and (16b).  We
can now \e{combine} the purely dynamical and transverse versions
of the Maxwell and Faraday laws which are given by Eqs.~(17a) and
(17b) into a \e{single} complex-valued equation by adding Eq.~(17b)
to Eq.~(17a) multiplied through by the imaginary unit $i$.  This is
readily seen to produce the result,
\re{
i\partial(\vE_T(\vr, t) + i\vB(\vr, t))/\partial t =
c\del\x (\vE_T(\vr, t) + i\vB(\vr, t)) - i\vj_T(\vr, t).
}{17c}
If we instead \e{subtract} Eq.~(17b) from Eq.~(17a) multiplied 
through by the imaginary unit $i$, we obtain,
\re{
i\partial(\vE_T(\vr, t) - i\vB(\vr, t))/\partial t =
-c\del\x (\vE_T(\vr, t) - i\vB(\vr, t)) - i\vj_T(\vr, t),
}{17d}
The difference between Eqs.~(17c) and (17d) above bears a
strong similarity to the difference between Eqs.~(7a)
and (7c) or that between Eqs.~(7b) and (7d).  As in those
instances, it is readily shown that Eqs.~(17c) and (17d)
above are, in fact, \e{equivalent}.  We kept Eqs.~(7a) and
(7b) \e{in preference} to Eqs.~(7c) and (7d) because, when
the transverse current source terms were dropped, the former
two evidenced a manifestly \e{nonnegative} Hamiltonian
operator, which is \e{physically appropriate} for a solitary
free particle.  Here, unfortunately, \e{neither} of the
candidate Hamiltonian operators $(c\h\del\x)$ and $(-c\h\del\x)$
turns out to be \e{nonnegative}, and \e{both} have the \e{further
peculiarity} of being \e{odd parity operators}, which is
\e{unacceptable} for a solitary free photon Hamiltonian operator.
We shall, in fact, need to \e{meld} a \e{piece} from \e{each} of
these two Hamiltonian operators to one another in such a way as
\e{produces in configuration representation} the \e{physically
appropriate} Hamiltonian operator for the solitary free photon
which Eqs.~(7a) and (7b), sans their transverse current source
terms, \e{have already delivered to us in Fourier vector variable
representation as} $|c\h\vk|$.  But before we discuss this melding
of pieces from the two Hamiltonian operators, we need to look at
the details of the passage to the source-free situation, and we
\e{also} need to pass from the complex-valued transverse-vector
\e{electromagnetic field strengths} that are found in Eqs.~(17c)
and (17d) to complex-valued transverse-vector \e{probability
density amplitudes}, since it is \e{these} which befit the quantum
description of a solitary particle (the transverse-vector character
of these probability density amplitudes reflects the the fact that
the particle has \e{only two} polarization degrees of freedom).  
In the source-free case that $\vj(\vr, t) = \vz$ and
$\rho(\vr, t) = 0$, it follows that $\phi(\vr, t) = 0$, and
therefore also $\vE_L(\vr, t) = \vz$, which, in turn, causes
$\vE_T(\vr, t)$ to be equal to $\vE(\vr, t)$.  Thus, after they
are multiplied through by $\h$, Eqs.~(17c) and (17d) become the
homogeneous time-dependent Schr\"{o}dinger-like equations,
\re{
i\h\partial(\vE(\vr, t) + i\vB(\vr, t))/\partial t =
c\h\del\x (\vE(\vr, t) + i\vB(\vr, t)),
}{17e}
and,
\re{
i\h\partial(\vE(\vr, t) - i\vB(\vr, t))/\partial t =
-c\h\del\x (\vE(\vr, t) - i\vB(\vr, t)),
}{17f}
whose respective ``wave functions'' $(\vE(\vr, t) \pm i\vB(\vr, t))$
are complex-valued vectors that are \e{purely transverse}, since,
in the source-free case, $\del\dt\vE = 0$ from the Coulomb law, as
well as $\del\dt\vB = 0$ from the Gauss law.  Their respective
``Hamiltonian operators'' $(\pm c\h\del\x)$ are \e{Hermitian} on
such a ``wave function'' space, since it is readily shown, using
integration by parts that,
\re{
{\ty\int}d^3\vr\,\lf[(\bv{a}(\vr, t))^\ast\dt(\pm\del\x\bv{b}(\vr, t))\rt] =
{\ty\int}d^3\vr\,\lf[(\pm\del\x\bv{a}(\vr, t))^\ast\dt\bv{b}(\vr, t)\rt].
}{17g}
\indent
Because the total purely electromagnetic field energy, which
is $\hf\int d^3\vr\,(|\vE(\vr, t)|^2 + |\vB(\vr, t)|^2)$, is
readily shown to be \e{independent of time} in the source-free
case (indeed this follows from the real and imaginary parts
of either of Eqs.~(17e) or (17f), in conjunction with Eq.~(17g)),
we can define \e{two} candidate solitary-particle complex-valued
vector wave functions which are \e{properly normalized to unity},
namely,
\re{\vPsi(\vr, t)\eqdf (\vE(\vr, t) + i\vB(\vr, t))/
({\ty\int}d^3\vr'\,(|\vE(\vr', t)|^2 + |\vB(\vr', t)|^2))^{\hf},
}{18a}
and \e{also} $(\vPsi(\vr, t))^\ast$, the \e{complex conjugate}
of this $\vPsi(\vr, t)$.  It is clear from Eq.~(18a) that,
\re{
{\ty\int}d^3\vr\,\lf[(\vPsi(\vr, t))^\ast\dt\vPsi(\vr, t)\rt] = 1,
}{18b}
a property which $(\vPsi(\vr, t))^\ast$ obviously shares,
\re{
{\ty\int}d^3\vr\,\lf[\vPsi(\vr, t)\dt(\vPsi(\vr, t))^\ast\rt] = 1.
}{18c}
We further note that, because in the source-free case
$\del\dt\vE(\vr, t) = 0$ and $\del\dt\vB(\vr, t) = 0$,
both $\vPsi(\vr, t)$ and $(\vPsi(\vr, t))^\ast$ are
\e{strictly transverse}, i.e.,
\re{
\del\dt\vPsi(\vr, t) = 0\ \m{and}\ \del\dt(\vPsi(\vr, t))^\ast = 0,
}{18d}
but that, as we see from Eqs.~(17e) and (17f), the time-dependent
Schr\"{o}dinger equations which $\vPsi(\vr, t)$ and
$(\vPsi(\vr, t))^\ast$ obey have Hamiltonian operators of opposite
sign,
\re{
i\h\dot\vPsi(\vr, t) = c\h\del\x\vPsi(\vr, t),
}{18e}
\m{}
\re{
i\h(\dot\vPsi(\vr, t))^\ast = -c\h\del\x(\vPsi(\vr, t))^\ast.
}{18f}
Unfortunately, \e{neither} of the Hamiltonian operators $(c\h\del\x)$
and $(-c\h\del\x)$ in the two Schr\"{o}dinger equations just given is
\e{nonnegative}, which, in light of the \e{nonnegativity of the purely
electromagnetic field energy}
$\hf\int d^3\vr\,(|\vE(\vr, t)|^2 + |\vB(\vr, t)|^2)$,
is an \e{absolutely necessary requirement} for the physically proper
description of a solitary free photon.  In \e{addition}, both of
these Hamiltonian operators are \e{odd parity operators}, which
implies that \e{reversing the parity} of any of their energy eigenstates
produces an energy eigenstate whose eigenenergy has the \e{opposite
sign} to that of the eigenenergy of the \e{original} energy eigenstate!
This \e{complete energy eigenspectrum sign symmetry} is again
incompatible with the need for solitary free photons to be of strictly
nonnegative energy.  The nonconservation of parity for a free solitary
photon which such a Hamiltonian implies is as well \e{incompatible with
the conservation of parity on the part of electromagnetic theory}.

In order to overcome these problems, we must \e{meld} the \e{nonnegative
part} of the eigenenergy/eigenstate spectrum of $(c\h\del\x)$ with the
\e{complementary positive part} of the eigenenergy/eigenstate spectrum 
of $(-c\h\del\x)$.  The resulting \e{nonnegative Hamiltonian operator}
will clearly be the \e{absolute value} of the operator $(c\h\del\x)$,
i.e., the operator $|c\h\del\x|$.  To gain further insight into the
precise nature of this \e{melded} Hamiltonian operator $|c\h\del\x|$,
let us look at the \e{purely transverse} eigenstates of $(c\h\del\x)$.
Each of these eigenstates is clearly a complex-valued transverse vector
field which consists of a \e{single Fourier component} whose Fourier
vector variable \e{direction} is \e{strictly orthogonal to the
longitudinal unit vector} $\vu_L(\vk)$.  Upon operating on such a
single Fourier component proportional to $e^{i\vk\dt\vr}$, the operator
$(\del\x)$ clearly becomes $i(\vk\x) = i|\vk|(\vu_L(\vk)\x)$, whereas
the \e{two-dimensional transverse Fourier vector variable space} is
clearly \e{fully spanned} by the \e{two transverse unit vectors}
$\vu_1(\vk)$ and $\vu_2(\vk)$, or, \e{much more usefully in this
particular instance}, the \e{two complex-valued transverse unit
vectors} $(2)^{-\hf}(\vu_1(\vk) \pm i\vu_2(\vk))$.  By making use
of the cross product identities of Eq.~(5d) in conjunction with
the facts just discussed, one readily verifies that the \e{purely
transverse} eigenstates of the Hermitian operator $(\del\x)$ are
of the form,
\re{
\vPsi^{(\pm)}_\vk(\vr)\eqdf (\vu_1(\vk) \pm i\vu_2(\vk))g^{(\pm)}(\vk)
 e^{i\vk\dt\vr},
}{19}
and have the \e{corresponding eigenvalues} $\pm |\vk|$ for the
operator $(\del\x)$.  Therefore the corresponding \e{two}
eigenvalues of the \e{melded} Hamiltonian operator $|c\h\del\x|$
on the above two \e{purely transverse} eigenstates are \e{both}
simply $|c\h\vk|$, in \e{complete agreement} with those of the
fully diagonalized Hamiltonian for the \e{two} time-dependent
Schr\"{o}dinger equations that are given by Eqs.~(11a) and (11b).
Thus we clearly see that an \e{equivalent} way to express the
\e{action} of our \e{melded} configuration-space Hamiltonian
operator $|c\h\del\x|$ on the \e{purely transverse subspace} is
as \e{simply} $|c\h\q{\vk}|$, which can be loosely styled as
$|c(-i\h\del)|$ in configuration representation.  Because of the
\e{purely transverse, two-dimensional nature} of the \e{subspace}
on which it is \e{obliged to operate}, the  nonnegative \e{melded}
configuration-space Hamiltonian $|c\h\del\x|$ \e{loses all trace
of its curl character} and becomes indistinguishable from the
loosely styled $|c(-i\h\del)|$, thus falling into line with the
$m=0$ case of the Eq.~(1) Hamiltonian, namely $|c\qvp| = 
|c\h\q{\vk}| = |c(-i\h\del)|$.

Summarizing, we have that the solitary free-photon wave function
in configuration representation is a complex-valued vector field
$\vPsi(\vr, t)$ which is \e{strictly transverse}, i.e.,
\re{
\del\dt\vPsi(\vr, t) = 0,
}{20a}
generally normalizable to unity (unless idealized),
\re{
{\ty\int}d^3\vr\,\lf[(\vPsi(\vr, t))^\ast\dt\vPsi(\vr, t)\rt] = 1,
}{20b}
and satisfies the time-dependent Schr\"{o}dinger
equation,
\re{
i\h\dot\vPsi(\vr, t) = |c(-i\h\del)|\vPsi(\vr, t),
}{20c}
in accord with the $m = 0$ instance of Eq.~(1).  The
loosely styled configuration-space Hamiltonian operator
$|c(-i\h\del)|$ of the above Schr\"{o}dinger equation
is \e{actually}, in configuration representation, a
symmetric nonlocal \e{integral operator} (something
which Klein, Gordon, Schr\"{o}dinger and Dirac sought
\e{desperately} to avoid, thereby playing \e{havoc}
with physical cogency), whose kernel is given by,
\re{
\la\vr|\,|c(-i\h\del)|\,|\vr'\ra =
\la\vr|\,|c\h\q{\vk}|\,|\vr'\ra =
-(c\h/(2\pi^2R))d^2(R/(R^2 + \ep^2))/dR^2,
}{21}
where $R\eqdf |\vr - \vr'|$ and $\ep$ is
an infinitesimal length.

\subsection*{Conclusion}

Finally, it is to be noted that the very first quantum theorist
was James Clerk Maxwell.  His celebrated equations faithfully
encompassed the correct quantum description of the solitary free
photon long before Erwin Schr\"{o}dinger was to accomplish the
same feat for the solitary \e{nonrelativistic} free particle.  And
Maxwell's formidable theoretical physics machinery \e{already}
yielded up the first instance of \e{intrinsic} particle degrees
of freedom, with all their subtlety!  By the grace of almost
cosmic coincidence, Maxwell, unlike any of his quantum theory
successors, could accomplish all this with no use whatsoever
of Planck's world-changing constant, which, still undiscovered,
silently awaited the future---the massless nature of the photon
permits Maxwell's magnificent equations to simply slip away from
$\h$'s grasp.

\end{document}